\documentclass[a4paper,11pt]{article}
\usepackage{pos}

\usepackage{graphicx}
\usepackage{xcolor}

\title{A search for Elves in Mini-EUSO data using CNN-based one-class classifier}
\ShortTitle{A search for Elves in Mini-EUSO data using CNN-based one-class classifier}

\author*[a]{Lech Wiktor Piotrowski}

\affiliation[a]{Faculty of Physics, University of Warsaw,\\
Pasteura 5, 02-093 Warsaw, Poland}

\onbehalf{for the JEM-EUSO Collaboration}
\emailAdd{science@lwp.email}

\abstract{
Mini-EUSO is a small, near-UV telescope observing the Earth and its atmosphere from the International Space Station. The time resolution of 2.5 microseconds and the instantaneous ground coverage of about $320\times 320$ km$^2$ allows it to detect some Transient Luminous Events, including Elves. Elves, with their almost circular shape and a radius expanding in time form cone-like structures in space-time, which are usually easy to be recognised by the eye, but not simple to filter out from the myriad of other events, many of them not yet categorised. In this work, we present a fast and efficient approach for detecting Elves in the data using a 3D CNN-based one-class classifier. 
}

\ConferenceLogo{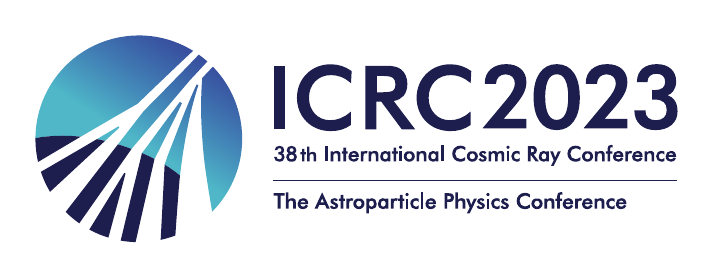}

\FullConference{%
38th International Cosmic Ray Conference (ICRC2023)\\
  26 July - 3 August, 2023\\
  Nagoya, Japan}

\begin{document}

\maketitle

\section{Introduction}
\label{sec:Introduction}

Mini-EUSO \cite{Bacholle_2021} is a small orbital telescope, designed within the JEM-EUSO programme \cite{bib:JEMEUSO}, observing the night-time Earth from the International Space Station (ISS) through a UV-transparent window inside the Zvezda module. It is composed of two 25 cm diameter Fresnel lenses focusing light on a Photon Detection Module (PDM) consisting of 36 multi-anode photomultipliers (MAPMTs), encompassing 2304 pixels. The field of view (FoV) is $44^{\circ}\times 44^{\circ}$, with a single-pixel side covering roughly 6 km on the ground, and the whole PDM more than 300 km. The spectral acceptance spans between 290 and 430 nm, making Mini-EUSO mostly a UV telescope. The PDM data is gathered in 3 time resolutions. D1 data consists of packets of 128 frames, each with 2.5 $\mathrm{\mu s}$ exposure, stored upon receiving a fast-events trigger from the FPGA. D2 data packet consists of 128 frames, each being an average of 128 D1 frames, forming a 320 $\mathrm{\mu s}$ block. It is collected after receiving a separate, slow-events trigger. D3 data are untriggered, forming a continuous ``movie'' with a single frame being an average of $128\times 128$ D1 frames, spanning 40.96 ms.
	
The PDM is a very sensitive instrument and thus can be damaged by excessive light. Thus, two main levels of protection were introduced. The first one switches a part (an ``EC-unit'' composed of 4 MAPMTs) of the detector to lower gain if a few very bright pixels are detected in it. This happens quite often when going over the cities, etc. The second one is an analogue over-current protection, sensitive to the summed signal in all the EC-unit pixels. The telescope is also equipped with a small near infra-red camera and a visible light camera set to take photos with 5 s exposure time, photodiodes for detecting the night/day transitions, and a small silicon photomultiplier.

Three time resolutions of Mini-EUSO allow it to observe a wide range of phenomena. It can create a near-UV map of Earth, observe vast numbers of meteors, and detect very fast atmospheric events, including Transient Luminous Events such as ELVES.

\section{ELVESs}
\label{sec:elves}

Transient Luminous Events (TLEs) are electrical discharges in the upper atmosphere, usually associated to thunderstorms. Their existence was predicted in 1920s by C. T. R. Wilson \cite{Wilson_1924}, and the first observation was performed by R. C. Franz in 1989 \cite{Franz1990TelevisionIO}. One of the types of TLEs are ELVESs (Emission of Light and Very low frequency perturbations due to Electromagnetic pulse Sources), discovered by Mesoscale Lightning Experiment in 1990. They appear as thick rings of light - a horizontal doughnut cross-section -- propagating through the ionosphere at an altitude of about 100 km, and are caused by an electromagnetic pulse from an underlying thunderstorm. They can be multiringed due to, mainly, reflection of the pulse from the ground, and the diameter goes to a few hundred kilometres. The ring expansion speeds lie around the speed of light, thus the whole phenomenon lasts typically about one millisecond.

Mini-EUSO was not designed for ELVESs observations, however its $\sim 200\times 200$ km FoV at the altitude of ionosphere, and the trigger at D1 2.5 $\mu$s frame length allows for registering these phenomena in a slightly different way than dedicated experiments. The telescope usually sees only part of the ring at the later stages of expansion, for at the beginning either the ring itself or accompanying light emission from the centre are too bright for our telescope. The observed ring thickness is of an order of a few pixels, changing during the propagation, along with the changing intensity. In 3D space consisting of 2 spatial x, y dimensions and 1 temporal t dimension, an ELVES appears as a part of an approximate, thick cone surface, or multi-cone surface with a common top in case of a multi-ringed ELVES. However, the propagation of the rings is often followed by a significant brightening of the central source, which causes lightning protection to be switched on for influenced parts of the detector, and thus the rings being followed or propagating through areas of lower sensitivity, making the observational picture more complicated.

There are a few common phenomena that may resemble ELVESs in our telescope, due to its design. First, any significantly brightening, non-diffuse source may look like an expanding doughnut in Mini-EUSO. This is due to saturation -- extendable dead time that causes reduction and then stop of photon counting for photons coming too close in time. The number of counts vs. light intensity dependence grows, reaching its maximum between 100 and 200 counts, then drops, reaching 0. Thus, any sufficiently bright source is visible as a doughnut, with a 0-counts centre, then counts sharply growing and slowly dropping with the distance from the centre, due to first exiting from the saturation and then going through the arms of the point spread function (PSF). If the intensity of the central source increases, the ``dead'' central area and the visible ``doughnut'' radius grow in time, similarly to an ELVES. The difference lies in the profile of the doughnut. The external slope of an ELVES is steeper, as it is caused primarily by the physical boundaries of the light emission, not the PSF. The internal slope of the ELVES is not as sharp, as the drop in light intensity is not caused by saturation. At least part of the ELVES interior should be at roughly a background level. Still, as mentioned before, an ELVES is often followed by the brightening of the central source and saturation, making the distinction more complicated.

The nature of the optics of the experiment results in ring-like structures in the PSF far from its centre. For the observable central source intensities, these are too dim to be observed, but for a brightening source outside the FoV they are visible as brightening, thick circles, often causing an illusion of movement. Also, the light protection switching in Mini-EUSO may result in some artificial structures slightly resembling an ELVES. It is important to note, that these events are caused by excessive light, which is common during thunderstorms -- conditions required to register an ELVES.

\section{The machine-learning based ELVESs searching algorithm}

The simple pixel over threshold search method is not sufficient for the ELVESs identification purpose, for the diffuse nature of the rings and the fact, that the hardware may trigger on something else than the rings themselves. In the ideal case, searching for a cone-like structure in x, y, t data packets is simple and gives much better results. Initially, we have attempted identification with a series of negative cuts and a fit of conical surface to the remaining cases. This method was, however, mainly troubled by outliers -- difficulties in selecting pixels belonging to the ELVES, especially at the late stages or for weak elve. More stable results were achieved when we were analysing each frame separately in circular Hough space and then estimated rings centres and radii in linear Hough space. However, the deviations from conicality especially for very weak events, reduce identification efficiency. The efficiency to background ratio is further decreased by the existence of ELVES-resembling background that may be seen independently but may also accompany an ELVES. Still, these factors do not have a significant influence on manual identification. This led us to the creation of a Machine Learning algorithm in the hope it can spot visual clues similarily that are easy to spot for humans, but complex to contain in a conventional data analysis method.

\subsection{The neural network architecture}

The problem of ELVESs identification in the Mini-EUSO data is a problem of so-called one-class classification. A packet of the data needs to be classified as containing an ELVES or not. While formally this would belong to the family of binary classifications, it is not in the sense most commonly used in Machine Learning based on pattern recognition. This problem differs significantly from categorising events into several (in this case two) defined classes, such as tutorial examples of recognising which of 9 digits a handwritten character represents, or if the analysed photo is a photo of a cat or a dog. In multi-class categorisation the neural network usually learns the shared characteristics of each class. Then, oversimplifying. The trained network estimates how each class's characteristic describes the given sample. In the ELVESs identification case, only one class is well-defined in terms that it shares a common characteristic -- the class representing the ELVESs. The events belonging to the second class -- ``not-ELVESs'' -- do not have to have anything in common apart from not being an ELVES. Thus, in a one-class classificator we need to force the network to learn the common characteristic of one class, but prevent it from attempting to find common features of the samples of the second class.

\begin{figure}[t]
		\begin{center}
            \includegraphics[width=0.49\textwidth]{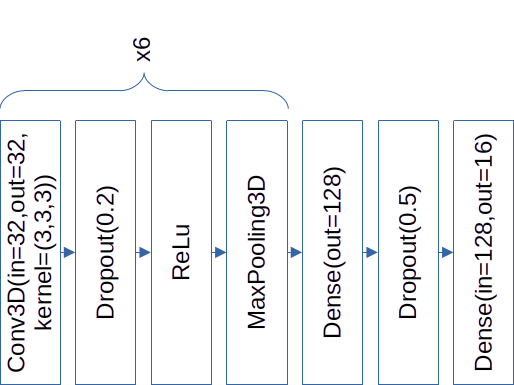}
    		\caption{The architecture of the network presented in this paper.}
			\label{fig:cnn_arch}
		\end{center}
\end{figure}

The implemented idea is based on \cite{Zhang2017TheAO}. We use a simple neural network based on 6 3D convolutional layers coded in PyTorch \cite{NEURIPS2019_9015}, with details shown on fig. \ref{fig:cnn_arch}. The network accepts a standard D1 Mini-EUSO packet of $48\times 48 \times 128$ values, where the first two dimensions denote x and y pixels, and the third number of frames. The network transforms the packet to 16 values, which can be interpreted as coordinates of the packet in a 16-dimensional space. In our case, if the Euclidean distance $D$ of the 16-dimensional point from 0 is lower than $m=2$, the packet is classified as an ELVES, and if higher it is classified as non-ELVES.

The loss function used for training:

$$\frac{1}{2}(Y\cdot D^2 + (1-Y)\cdot (\mathrm{max}(0, m-D))^2) $$

where $Y=1$ for ELVESs and $Y=0$ for non ELVESs, approaches 0 for ELVESs position approaching 0 in the 16-dimensional space, while for non-ELVESs it is 0 if their distance from 0 is bigger than 2, and grows if the non-ELVESs approach 0 within the sphere of radius 2.

\subsection{The training and validation set}

The amount of ELVESs in Mini-EUSO data is very limited. Initially, we were operating on a set of 19 events detected with other methods, where 11 were used in the training process and 8 in external validation. To increase the number of events, the set was augmented by spatial flips and $90^{\circ}$ rotations, and finally by introducing random fluctuations of pixel values to the events. False events set was generated from packets close in time, but not belonging to an ELVES, as these packets were well inspected. Later, packets with specific types of events frequently misclassified as ELVESs were added to the set. The process resulted in roughly 1200 events of both classes, where 70\% was used for training and 30\% for validation. These are not big numbers in the world of Machine Learning, but one has to keep in mind that they were employed for a one-class classifier recognising a rather clear pattern. The remaining 8 elves were augmented in the same way and accompanied by false events not included in the internal training/validation set, and used for external validation to assess the neural network model generalisation capabilities.

All the data were flat-fielded\footnote{Flat-fielding is a process of uniformising the detector's response by dividing the data by calibration data obtained with a uniformly illuminated instrument.}, then "Gaussianised" with Anscombe transformation. Finally, extreme values were clipped, and each pixel had its mean value subtracted and was divided by its standard deviation.

\section{Results}

\begin{figure}[t]
		\begin{center}
            \includegraphics[width=0.49\textwidth]{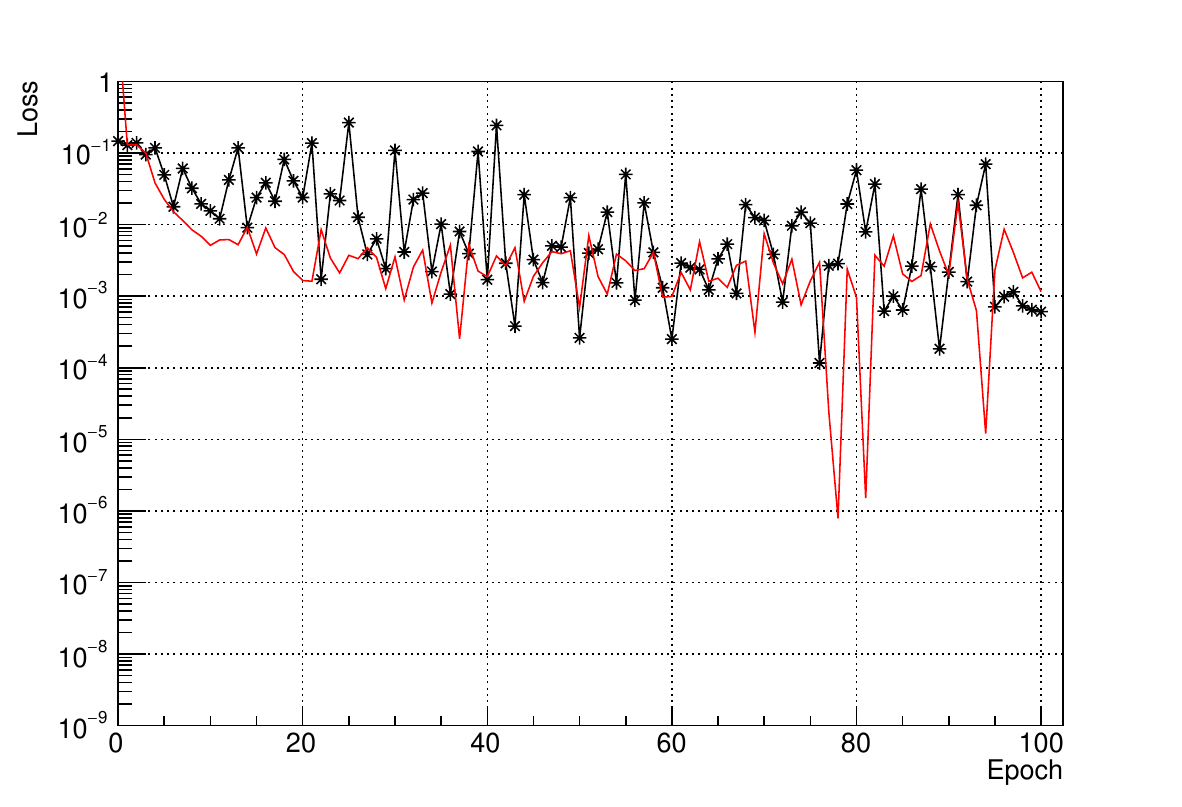}
			\includegraphics[width=0.49\textwidth]{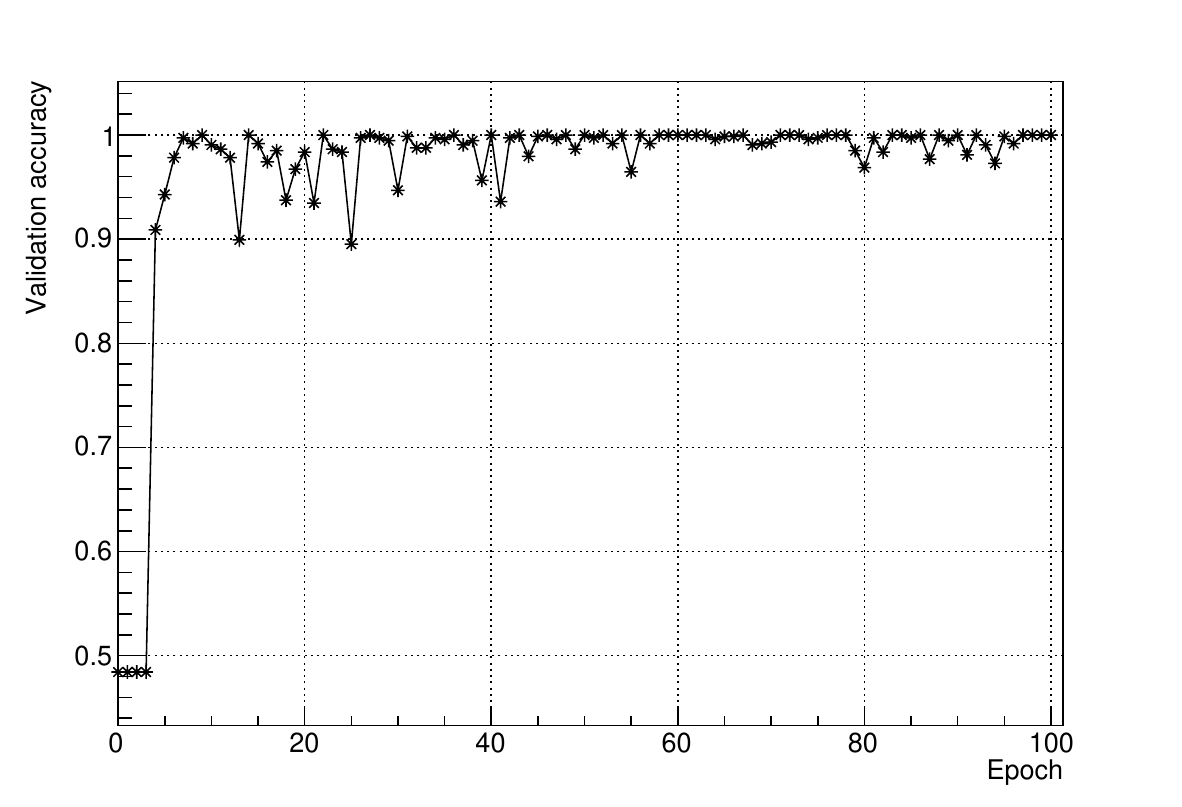}
			\caption{{\em Left:} The loss of the training set (red) and internal validation set (black). {\em Right:} The accuracy of the internal validation set.}
			\label{fig:loss_acc}
		\end{center}
\end{figure}

The training during the development of the network architecture was tested with three batch sizes: 8, 16 and 32 packets, the last one being close to the maximum that was supported on the used NVIDIA GeForce RTX 2060 GPU with 6 GB of RAM. However, it was quickly discovered that the best results are almost always obtained with the batch size 8. The drawback is high instability of the loss and accuracy, as can be seen on the fig. \ref{fig:loss_acc}, depicting training progress in our final design. The network was usually reaching close to 100\% accuracy within the first 15 epochs of training, while the loss reduction was most significant within the first 50 epochs. The shown loss and accuracy curves for training and internal validation sets do not show obvious overfitting.

However, such a small initial data set can not be trusted even when augmented. Therefore, after the training, we were checking the efficiency of the snapshots of the network for chosen epochs on data files of all the ELVESs found so far. With the further availability of data, this set has grown to 29 ELVESs spanning through 35 packets. Depending on the network hyperparameters and the epoch of training, the best efficiencies were varying between 80\% and 95\%, and the amount of misidentified background packets between 0.5\% and 2\% out of 1554 real packets analysed in the external validation. While these results may seem mediocre, they are already better than our conventional algorithms in terms of efficiency and far better in terms of non-ELVESs classification, and provide some knowledge about the network generalisation capability.

The knowledge that the model generalises at least to some degree was crucial for the final step, which was training the model, giving the best result on the full set of detected ELVESs. Given a very small initial data set, this have a potential of increasing the networks' efficiency in recognising ELVESs, but at the same time prevents external validation, and we run a risk of unnoticed overfitting. Therefore, the training was performed for just one epoch. The final model was run on all the available Mini-EUSO data. It was able to properly identify all the ELVESs-containing packets and found 8 new ELVESs. At the same time, it misidentified 308 packets as ELVESs in the whole data set containing hundreds of thousands of packets.

\section{Summary}
\label{sec:summary}

The main purpose of this work was to create a machine learning based algorithm that would be more efficient in identifying ELVESs in the data than conventional algorithms prepared earlier. This task was completed successfully with the model being 100\% efficient on the pre-detected ELVESs, misidentifying less tan 0.1\% of packets and capable of detecting new ELVESs. Assuming that the final model's efficiency is not worse than the efficiency of a model trained on the limited data set, we expect the full efficiency to be at least 80\% and very likely much better than that. Unfortunately this can remain only an educated guess, because a set of ELVESs detected with alternative, similarly efficient algorithm does not exist, nor does a simulated data set.

Until we find that the algorithm is not able to identify some ELVESs identified with other methods, improving it is technically difficult. We could try to increase spatial separation of the classes in their resulting 16 dimensions, but that most likely requires creating an alternative loss function. The understanding of the model's limits could be improved with analysing the efficiency vs some effective signal-to-noise ratio. However, even the standard signal-to-noise ratio is not trivial to estimate for ELVESs which are diffuse rings of increasing radius, even more difficult to modify, and it would be just a part of an effective signal-to-noise ratio, that must include also the light background conditions, light-protection response of the detector, etc. Still, this kind of estimation would be useful also for other purposes and could help improve the model design and parameters, and we intend to prepare it in the future.

\bibliographystyle{JHEP}
\bibliography{lwp_icrc_2023}{}

\providecommand{\href}[2]{#2}\begingroup\raggedright\begin{thebibliography}{1}

\bibitem{Bacholle_2021}
S.~Bacholle, P.~Barrillon, M.~Battisti, A.~Belov, M.~Bertaina, F.~Bisconti
  et~al., \emph{Mini-euso mission to study earth uv emissions on board the
  iss}, \href{https://doi.org/10.3847/1538-4365/abd93d}{\emph{The Astrophysical
  Journal Supplement Series} {\bfseries 253} (2021) 36}.

\bibitem{bib:JEMEUSO}
Y.~Takahashi, J.-E.~Collaboration et~al., \emph{{The JEM-EUSO Mission}},
  {\emph{New Journal of Physics} {\bfseries 11} (2009) 065009}.

\bibitem{Wilson_1924}
C.T.R.~Wilson, \emph{The electric field of a thundercloud and some of its
  effects}, \href{https://doi.org/10.1088/1478-7814/37/1/314}{\emph{Proceedings
  of the Physical Society of London} {\bfseries 37} (1924) 32D}.

\bibitem{Franz1990TelevisionIO}
R.~Franz, R.J.~Nemzek and J.R.~Winckler, \emph{Television image of a large
  upward electrical discharge above a thunderstorm system}, {\emph{Science}
  {\bfseries 249} (1990) 48 }.

\bibitem{Zhang2017TheAO}
M.~Zhang, J.~Wu, H.~Lin, P.~Yuan and Y.~Song, \emph{The application of
  one-class classifier based on cnn in image defect detection}, {\emph{Procedia
  Computer Science} {\bfseries 114} (2017) 341}.

\bibitem{NEURIPS2019_9015}
A.~Paszke, S.~Gross, F.~Massa, A.~Lerer, J.~Bradbury, G.~Chanan et~al.,
  \emph{Pytorch: An imperative style, high-performance deep learning library},
  in \emph{Advances in Neural Information Processing Systems 32},
  pp.~8024--8035, Curran Associates, Inc. (2019),
  \href{http://papers.neurips.cc/paper/9015-pytorch-an-imperative-style-high-performance-deep-learning-library.pdf}{http://papers.neurips.cc/paper/9015-pytorch-an-imperative-style-high-performance-deep-learning-library.pdf}.

\end{thebibliography}\endgroup

\newpage
{\Large\bf Full Authors list: The JEM-EUSO Collaboration\\}

\begin{sloppypar}
{\small \noindent
S.~Abe$^{ff}$, 
J.H.~Adams Jr.$^{ld}$, 
D.~Allard$^{cb}$,
P.~Alldredge$^{ld}$,
R.~Aloisio$^{ep}$,
L.~Anchordoqui$^{le}$,
A.~Anzalone$^{ed,eh}$, 
E.~Arnone$^{ek,el}$,
M.~Bagheri$^{lh}$,
B.~Baret$^{cb}$
% ...
}
\end{sloppypar}
\vspace*{.3cm}

{ \footnotesize
\noindent
% ...
$^{cb}$ Universit\'e de Paris, CNRS, AstroParticule et Cosmologie, F-75013 Paris, France\\
% ...
$^{ed}$ Istituto Nazionale di Fisica Nucleare - Sezione di Catania, Italy\\
% ...
$^{eh}$ INAF - Istituto di Astrofisica Spaziale e Fisica Cosmica di Palermo, Italy\\
% ...
$^{ek}$ Istituto Nazionale di Fisica Nucleare - Sezione di Torino, Italy\\
$^{el}$ Dipartimento di Fisica, Universit\`a di Torino, Italy\\
% ...
$^{ep}$ Gran Sasso Science Institute, L'Aquila, Italy\\
% ...
$^{ff}$ Nihon University Chiyoda, Tokyo, Japan\\ 
% ...
$^{ld}$ University of Alabama in Huntsville, Huntsville, AL, USA\\
$^{le}$ Lehman College, City University of New York (CUNY), NY, USA\\
% ...
$^{lh}$ Georgia Institute of Technology, USA\\
% ...
}

\end{document}